\documentclass[10pt,twocolumn,twoside]{IEEEtran}

\usepackage{graphicx}
\usepackage{booktabs}
\usepackage{subfigure}
\usepackage{setspace}
\usepackage{multicol}
\usepackage{multirow}
\usepackage{epstopdf}
\usepackage{amsmath}
\usepackage{bm}
\usepackage{cite}
\usepackage{subfigure}
\usepackage{amssymb}
\usepackage{gensymb}
\usepackage{amsfonts}
\usepackage{mathrsfs}
\usepackage{amsmath}
\usepackage{algorithm}
\usepackage{algorithmic}
\usepackage{amsthm}

\usepackage{tabularx}
\usepackage{color}
\usepackage{balance}
\usepackage{mathrsfs}
\usepackage{setspace}
\usepackage{amsthm}
\usepackage{array}
\usepackage{cases} 
\newcommand{\PreserveBackslash}[1]{\let\temp=\\#1\let\\=\temp}
\newcolumntype{C}[1]{>{\PreserveBackslash\centering}p{#1}}
\usepackage[flushleft]{threeparttable}
\usepackage{setspace}
\newcommand{\tabincell}[2]{\begin{tabular}{@{}#1@{}}#2\end{tabular}}  
\usepackage{acronym}

\acrodef{snr}[SNR]{signal-to-noise ratio}%
\acrodef{ris}[RIS]{reconfigurable intelligent surface}
\acrodef{pin}[PIN]{positive-intrinsic-negative}
\acrodef{bs}[BS]{base station}
\acrodef{csi}[CSI]{channel state information}

\begin{document}

\bibliographystyle{IEEEtran} 

\title{Reconfigurable Intelligent Surfaces for 6G:  Nine Fundamental Issues and One Critical Problem}
\author{ 
	Zijian Zhang,~\IEEEmembership{Student Member,~IEEE}, and Linglong Dai,~\IEEEmembership{Fellow,~IEEE}\\
	\vspace{0.5em}
	{\it (Invited Paper)}
	\vspace{-2em}
	\thanks{This work was supported in part by the National Key Research and Development Program of China (Grant No. 2020YFB1805005), in part by the National Natural Science Foundation of China (Grant No. 62031019), and in part by the European Commission through the H2020-MSCA-ITN META WIRELESS Research Project under Grant 956256. {\it (Corresponding author: Linglong Dai.)}
	}
	\thanks{Zijian Zhang and Linglong Dai are with the Department of Electronic Engineering, Tsinghua University, Beijing 100084, China, and also with the Beijing National Research Center for Information Science and Technology (BNRist), Beijing 100084, China. (e-mails: zhangzj20@mails.tsinghua.edu.cn; daill@tsinghua.edu.cn).}
}
\maketitle
\IEEEpeerreviewmaketitle
\begin{abstract}	
Thanks to the recent advances in metamaterials, reconfigurable intelligent surface (RIS) has emerged as a promising technology for future 6G wireless communications. Benefiting from its high array gain, low cost, and low power consumption, RISs are expected to greatly enlarge signal coverage, improve system capacity, and increase energy efficiency. In this article, we systematically overview the emerging RIS technology with the focus on its key basics, nine fundamental issues, and one critical problem. Specifically, we first explain the RIS basics, including its working principles, hardware structures, and potential benefits for communications. Based on these basics, nine fundamental issues of RISs, such as ``What's the differences between RISs and massive MIMO?'' and ``Is RIS really intelligent?'', are explicitly addressed to elaborate its technical features, distinguish it from existing technologies, and clarify some misunderstandings in the literature. Then, one critical problem of RISs is revealed that, due to the ``multiplicative fading'' effect, existing passive RISs can hardly achieve visible performance gains in many communication scenarios with strong direct links. To address this critical problem, a potential solution called active RISs is introduced, and its effectiveness is demonstrated by numerical simulations.
\end{abstract}
\section{Introduction}
To meet the growing needs of high-quality services in future 6G communications, the increase of key performance indicators, such as spectrum efficiency and energy efficiency, has always been the research focus \cite{Zhengquan'19}. Recently, a new technology named {\it \ac{ris}} is emerging as a potential candidate for 6G, and it has attracted great attention in the communications community \cite{Basar'19,Wu'19,Huang'18'2,LinglongDai}. 
\par
As a new physical-layer technology, \acp{ris} were developed from metamaterials \cite{LinglongDai}. Different from the natural electromagnetic (EM) materials, the reconfigurable metamaterials can adaptively adjust the EM characteristics of incident waves according to the external stimulus. Utilizing this property, an RIS is usually designed as a reflective array equipped with a large number of passive elements each being able to reflect the incident signals passively with a controllable phase shift \cite{Wu'19}. By properly adjusting the phase shifts of massive elements, the reflection-type beamforming with high array gains can be achieved by \acp{ris} \cite{Huang'18'2}. Since their working mechanism is nearly passive without the need of radio frequency (RF) components, compared with the traditional massive MIMO, RISs have lower cost, lower power consumption, and much smaller noise \cite{Basar'19,Wu'19,Huang'18'2}. Benefiting from these advantages, \acp{ris} have recently attracted extensive research interests to enhance the wireless communications through enlarging signal coverage \cite{Basar'19}, improving channel capacity \cite{Wu'19}, and increasing energy efficiency \cite{Huang'18'2}.
\par
\begin{figure*}[!t]
	\centering
	\includegraphics[width=1\textwidth]{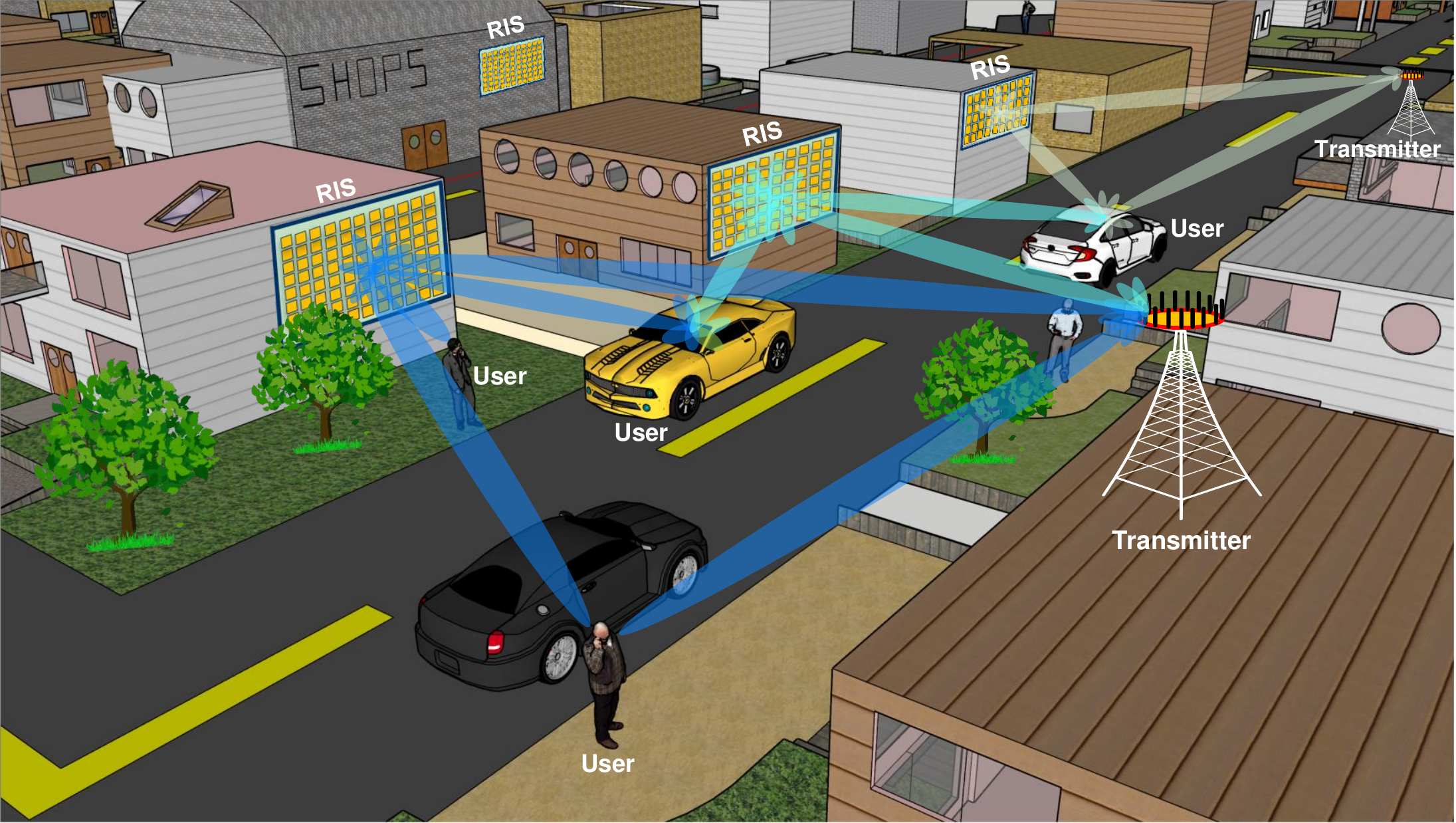}
	\caption{A typical communication scenario where RISs enhance wireless transmissions via passive beamforming.}
	\label{img:overview}
\end{figure*}
In this article, we systematically overview the RIS technology through explaining its key basics, clarifying nine fundamental issues, and revealing one critical problem. Specifically, the basics of RIS including its working principles, design methods, and potential benefits for communication systems, are explained at first. Based on these basics, nine fundamental issues of RISs, such as ``\textit{What's the differences between RIS and massive MIMO}?'', ``\textit{Is RIS just reflective}?'', and ``\textit{Is RIS really intelligent}?'', are explicitly addressed to elaborate their technical features, distinguish them from existing technologies, and clarify some misunderstandings in the literature. Then, we reveal one critical problem that, due to the ``\textit{multiplicative fading}'' effect, common RISs with hundreds of elements can hardly achieve visible gains in the communication scenarios with strong direct links. To address this critical problem, a potential solution called active RISs is introduced, and its effectiveness is demonstrated by experimental measurements and simulations. We expect that this article will inspire and stimulate more creative ideas and solutions to advance the application of RISs for future 6G wireless communications.

\section{RIS Basics}
In this section, we explain the basics of RISs. The working principle of RISs is explained at first. Then, two common hardware design methods of RISs are introduced. Finally, the potential benefits brought by RISs for wireless communications are discussed.

\subsection{Working Principle of RISs}
As the full name of RISs suggests, an RIS is a thin plate array composed of a large number of reconfigurable elements with phase-shift modules behind. Each RIS element works as a controllable scatterer, which can adjust some EM characteristics of the incident waves, such as amplitude, phase, polarization, and frequency. Although RISs have many anomalous EM properties, such as polarization adjustment and asymmetric reflection \cite{Basar'19}, most existing RIS-related works mainly focus on its phase-shifting function \cite{Basar'19,LinglongDai,Wu'19,Huang'18'2}, through which the passive beamforming at RISs can be realized. Specifically, by properly adjusting phase shifts of all RIS elements, the desired signals passively scattered by RISs can be coherently added at the receivers, while the undesired signals, such as multi-user interference, can be destructively combined to suppress their negative impacts \cite{Wu'19}. Fig. \ref{img:overview} intuitively illustrates a typical communication scenario where multiple RISs cooperatively work with several transmitters to enhance the transmissions via passive beamforming. 
\par
Thanks to its passive structure without the need of active components, RISs have low cost, low power, and negligible thermal noise \cite{Huang'18'2,Wu'19,Basar'19}. Particularly, due to the negligible thermal noise, the most essential benefit of RISs is the expected ``square-law'' asymptotic array gain \cite{Wu'19}. That is to say, for an RIS with $N$ elements, the receiver \ac{snr} achieved by passive beamforming is proportional to $N^2$, which is $N$ times larger than that can be achieved by the standard massive MIMO beamforming (proportional to $N$). Besides, the deploying RISs also efficiently increases the degrees of freedom (DoFs) of wireless channels to support multi-stream transmissions, especially for the sparse channels in millimeter-wave (mmWave) and terahertz (THz) communications. Benefiting from these advantages, RISs are expected to bring significant capacity gains for future 6G wireless communications.

\subsection{Hardware Design for RISs}\label{subsec:design}
Different from natural scatterers, RISs are expected to re-radiate the incident signals with some controllable EM characteristics. According to different reconfiguring mechanisms, RISs can be generally divided into two major categories, i.e., stimuli-responsive RISs and voltage-driven RISs \cite{Chen'16}. 
\par
\begin{figure}[!t]
	\centering
	\includegraphics[width=0.45\textwidth]{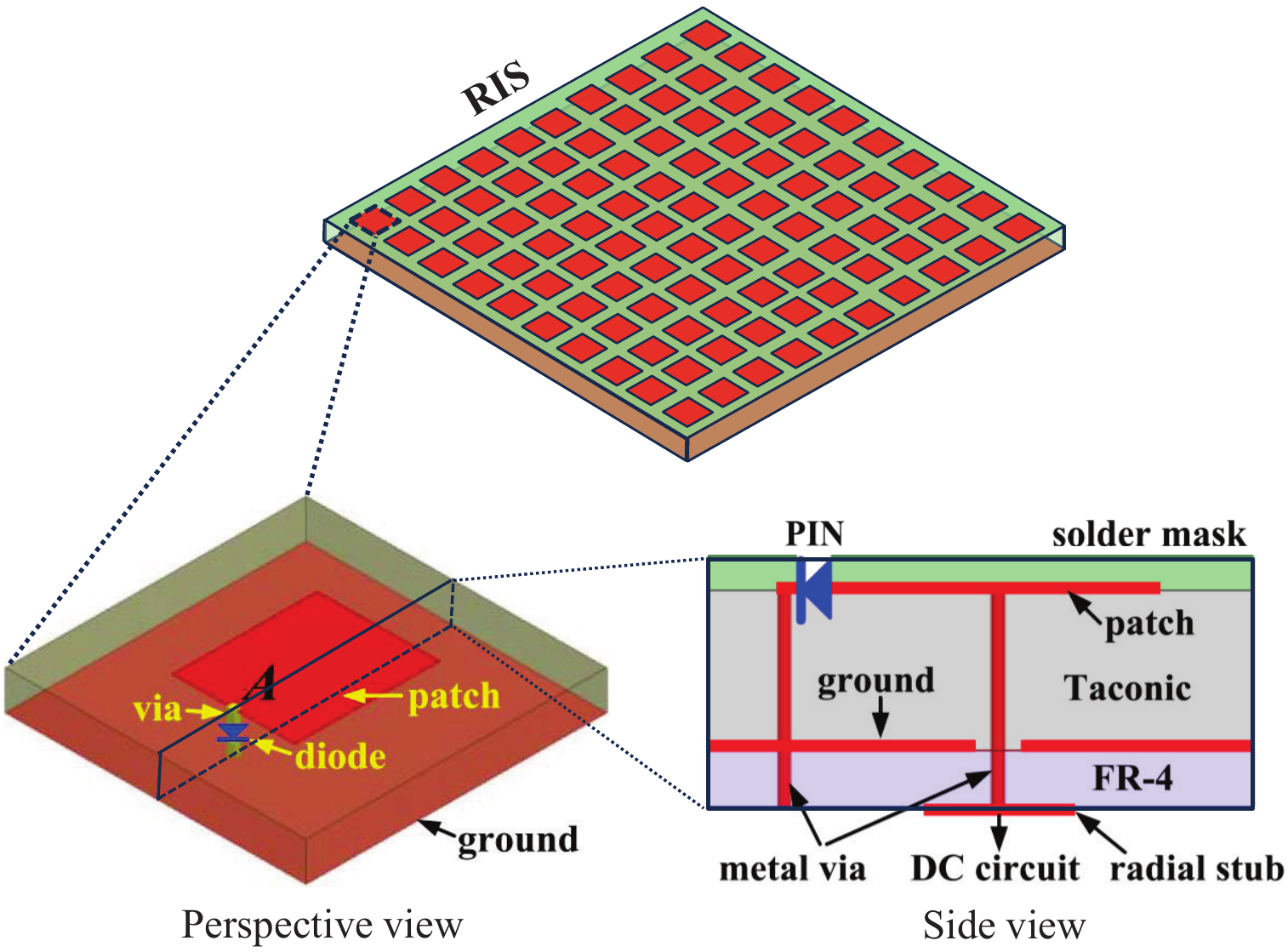}
	\caption{A schematic diagram of a typical phase-adjustable RIS \cite{Yang'17}.}
	\label{img:RISelement}
\end{figure}

For the former category, the typical stimuli-responsive RISs are usually made of synthetic reconfigurable materials, such as electric tuning materials, magnetic tuning materials, light tuning materials, etc \cite{Chen'16}. The physical properties of stimuli-responsive RISs will be reconfigured when RISs are controlled by different external stimuli, such as strong magnetic field and physical pressure, and thus the desired EM wave response can be achieved \cite{Chen'16}. However, due to the large size and high cost of stimuli generators, the stimuli-responsive RISs are not practical. 

As an alternative, the latter category of RISs, i.e., the voltage-driven RISs, is more common for communication systems \cite{LinglongDai}. Different from the stimuli-responsive RIS elements, the structure of voltage-driven RIS elements is equivalent to a patch antenna terminated with an tunable impedance. The impedance is usually realized by inductance-capacitance (LC) circuits and adjusted by the bias voltage, which controls the phase shift of each element. For example, as shown in Fig. \ref{img:RISelement}, the authors in \cite{Yang'17} adopted a typical switching-diode to control a voltage-driven RIS element, where the \ac{pin} diode is equivalent to a tunable LC circuit, and the phase shift can be controlled by the ON/OFF state the of \ac{pin} diode. Then, the whole RIS array is realized through the periodic arrangement of a large number of RIS elements. Thanks to this flexible design structure and operating mechanism, voltage-driven RISs usually have low cost and low power consumption for practical applications.
\subsection{Potential Benefits of RISs}\label{sub:adv}
By deploying RISs, a smart wireless environment can be established to enhance transmissions, where three major benefits for communications are expected \cite{Basar'19}. Firstly, RISs can enlarge the signal coverage in complex wireless environments \cite{Basar'19}. In the scenarios where the direct link from the transmitter to receiver is very weak due to blockage, RISs can be deployed to provide an additional reflected link and construct the wireless communications. Secondly, RISs can increase the spectrum efficiency \cite{Wu'19}. Particularly, RISs can strengthen the end-to-end link from the transmitter to receiver, and the noise introduced by RISs is negligible. Thereby, the desired signals from the transmitter can reach the receiver with high power without the additionally introduced noise \cite{Wu'19}, which can improve the channel capacity. Finally, RISs can increase the energy efficiency \cite{Huang'18'2}. Without the need of high-power RF components such as power amplifier, ADC/DAC, etc, RISs passively reflect the incident signals with low power consumption. Thus, an RIS is similar to a natural full-duplex relay but with nearly zero direct-current power consumption \cite{Yang'17}, which provides a promising solution for low-power communications \cite{Huang'18'2}.

\section{Nine Fundamental Issues of RISs}\label{sec:five}
Based on the RIS basics above, in this section, nine fundamental issues of RISs are explicitly addressed to elaborate its technical features, distinguish it from existing technologies, and clarify some misunderstandings in the literature.
\begin{table*}[t]
	\centering
	\normalsize
	\caption{Comparison of RIS with other related technologies.}
	\label{table:2}
	\vspace{0.8em}
	\begin{tabular}{|c|c|c|c|c|c|c|c|c|c|}
		\toprule  
		\hline
		{\bf Technology}&{\bf \tabincell{c}{Operating\\ Mechanism}}& {\bf Duplex}& {\bf Role} & {\bf \tabincell{c}{Signal\\ Detection}}&{\bf \tabincell{c}{Timeslot \\Synchronization
		}}&{\bf Cost}& {\bf \tabincell{c}{Power\\ Consumption}} \\ 
		\hline
		\tabincell{c}{Massive\\MIMO} &\tabincell{c}{active,\\ transmit/receive}&HD/FD&\tabincell{c}{source/\\ destination}&yes&yes&\tabincell{c}{very\\ high}&\tabincell{c}{very\\ high}\\
		\hline
		DF relay &\tabincell{c}{active,\\ decode-and-forward}&HD/FD&helper&yes&yes& \tabincell{c}{very\\ high}& \tabincell{c}{very\\ high}\\
		\hline
		AF relay &\tabincell{c}{active,\\ amplify-and-forward}&HD/FD&helper&no&yes& high& high\\	
		\hline
		Backscatter &passive, reflect&FD&{source}&no& no& low& low\\
		\hline
		Passive RIS&passive, reflect&FD&{helper}&no& no& low& low\\
		\hline  
		\hline
	\end{tabular}
\end{table*}
\subsection{Issue 1: What's The Difference Between RIS and Massive MIMO?}
RISs are usually considered to be similar to massive MIMO with more antennas/elements. For a clear comparison, here we summarize the key features of several related technologies in Table \ref{table:2}. It is true that, both massive MIMO and RIS use large-scale arrays to achieve high beamforming gains, while they have three essential differences. Firstly, massive MIMO is equipped with active RF components such as RF chains and signal processing modules, while RISs are composed of passive components such as capacitance and inductance. In this way, massive MIMO can detect and process baseband signals, while RISs cannot. This is the key reason why RISs are much cheaper and have much lower power consumption than massive MIMO. Secondly, massive MIMO works as an active source that generates and transmits the signals, or a destination that receives and reconstructs the signals, while RISs usually work as passive helpers that passively reflect the incident signals. Finally, different from the half-wavelength antenna spacing of massive MIMO, the spacing among RIS elements is usually much smaller than half wavelength. The reason is that, RISs only reflect without receiving signals, thus it is unnecessary to follow the spatial sampling theorem by using the half-wavelength antenna spacing. The smaller antenna spacing allows RISs to deploy more elements in a given aperture, so that it can maximize the radiation area and achieve high beamforming gains.

\subsection{Issue 2: What's The Difference Between RIS and Relay?}
RISs are also considered to be similar to the existing relays, since they both can work as helpers to enhance the wireless communications between the transmitter and the receiver. However, there are several major differences, as summarized in Table \ref{table:2}. Most important of all, decode-and-forward (DF) relays and amplify-and-forward (AF) relays are active, while RISs are passive. Equipped with high-cost and high-power RF components, DF and AF relays receive the incident signals and actively generate the forwarded signals \cite{Renzo'200}. By contrast, RISs just passively reflect the incident signals without RF components \cite{Basar'19}. Besides, DF relays can detect and decode the received signals thanks to the baseband signal procossing modules, while RISs do not have any baseband signal processing capability.
\par
Therefore, an RIS works similarly as a passive full duplex (FD) AF relay without signal amplification, while their operating mechanisms are different \cite{Renzo'200}. Specifically, the equipped RF components allow FD AF relays to simultaneously combines the received signals and precodes the forwarded signals, leading to a long latency. For an FD AF relay-aided system, the signals received by the receiver in one timeslot always carry two different symbols, one from the direct link and the other from the relay-aided link. To decode a symbol, two signals received in two adjacent timeslots are combined at the receiver \cite{Renzo'200}, which requires timing synchronization. By contrast, RISs cannot process the signals, and it just provides a reflection path with a nanosecond-scale latency \cite{Basar'19}, which allows the receiver to receive the signals carrying the same symbol from the direct link and RIS-aided link in one timeslot \cite{Renzo'200}.

\subsection{Issue 3: Is RIS Just Reflective?}
Most of existing works have regarded RISs as reflective arrays, which is also evidenced by another widely used name in the literature, i.e., intelligent reflecting surface (IRS). Actually, the reflective array is only one typical realization of RISs, and RISs can be designed in other forms. For example, a promising realization of RISs is the transmissive array \cite{Kunzan'22}, of which each RIS element can individually shift the phase of the signal passing through it. By deploying these transmissive RISs to replace the walls of the buildings, the wireless signals from outdoor can be precisely focused on the indoor users, which is a promising way to enhance communications and reduce the electromagnetic pollution. Besides, RISs can also be semi-reflective-transmissive \cite{Liu'21}. By adjusting the electromagnetic transmittance of each element, this kind of RISs allows some part of the incident signals to be reflected and the other part to pass through it \cite{Liu'21}. This structure enables RISs to serve the users indoor and outdoor simultaneously, which is expected to improve the space utilization of smart wireless environment \cite{Basar'21}.

\subsection{Issue 4: Is RIS Really Passive?}\label{subsec:RISpassive}
It is very common to find the statement ``RIS is passive'' in the literature \cite{Wu'19,LinglongDai,Basar'19,Huang'18'2}, since RISs passively shift the phases of the reflected signals without active RF components involved. However, actually this statement is not accurate. Although the signal process at RISs is passive due to their passive operating mechanism, RISs themselves are not completely passive due to the use of active components for driving and reconfiguring RISs. For example, a typical RIS element can be viewed as a patch antenna terminated with an tunable impedance. This tunable impedance can be realized by a \ac{pin} diode as shown in Fig. \ref{img:RISelement}, and two ON/OFF states of the \ac{pin} diode correspond to two different impedances for phase shifting. Then, to drive and reconfigure RISs, active circuits are required to generate the current to keep and switch the state of the \ac{pin} diode so as to hold and reset the desired phase shift. Particularly, the current to keep the \ac{pin} diode in the ON state is 10 mA at 1.4 V \cite{Yang'17}. In this way, the tunable impedance is actually connected to active components, which makes RISs become semi-passive. 

\subsection{Issue 5: Does RIS Not Introduce Noise?}\label{subsec:noise}
Thanks to the passive operating mechanism of RISs, the noise power introduced at RISs is much lower. Thereby, many existing works has reasonably assumed that the noise power at RIS is negligible \cite{Basar'19,LinglongDai,Wu'19,Huang'18'2}. However, RISs actually do introduce noise. Particularly, the radiation component of the RIS element is a patch antenna as shown in Fig. \ref{img:RISelement}. As long as the temperature is not at absolute zero, the Brownian motion of electrons naturally exists and determines the noise generation, which doesn't depend on whether RISs are passive or not \cite{Best'13}. According to the Nyquist noise theorem, the noise power introduced at each element is no lower than the theoretical bound, which is the product of the Boltzmann’s constant, the operating bandwidth, and the antenna noise temperature with a standard value of 290 K \cite{Best'13}. Furthermore, due to the noise factor of practical systems, the noise power introduced at each element will be higher in practice. For instance, from the experimental measurements in \cite[Fig. 14]{Kimionis'14}, one can note that a passive reflector working on a 250-kHz subcarrier is verified to introduce the noise power of about -107 dBm, which is much higher than its natural lower bound.
\par
To see the negative influence of noise on communication performances, here we consider an example. For a 10000-element RIS with 100 MHz operating bandwidth, the noise power at RIS can be estimated to be not lower than -54 dBm. If the path loss of the RIS-receiver link is -30 dB, the noise power at the receiver will be -84 dBm, which can compete with the original noise power at the receiver \cite{Wu'19}. In this case, when the number of RIS elements is large enough, the overall noise at the receiver will be gradually dominated by the noise introduced by RIS, which becomes non-negligible and will adversely influence the channel capacity.

\subsection{Issue 6: Can RIS Achieve The ``Square-Law'' Asymptotic Array Gain?}
In \cite{Wu'19}, an RIS is verified to achieve the ``square-law'' asymptotic array gain, which means that the receiver \ac{snr} achieved by the passive beamforming of RISs is asymptotically proportional to $N^2$ when the number of RIS elements $N$ is large. The reason behind this encouraging fact is that, the desired signal power at the receiver is proportional to $N^2$ thanks to the coherent superposition of received signals, while the undesired noise power at the receiver always remains unchanged due to the negligible noise assumption \cite{Zhang'22}.
\par
However, the ``square-law'' asymptotic array gain doesn't always hold in practice for two reasons. Firstly, according to the law of energy conservation, the receiver cannot receive the power higher than the transmit power. The receiver \ac{snr} is always bounded by the ratio of the transmit power to the noise power at the receiver, which doesn't depend on the number of RIS elements $N$. Secondly, beyond the ideal assumption of negligible noise introduced by RISs, the thermal noise introduced by RIS actually exists, as we have explained above. When the number of RIS elements $N$ is very large, the overall noise at the receiver will be dominated by the RIS noise. In this case, the overall noise power at the receiver will be proportional to $N$ instead of a constant. As a result, the receiver \ac{snr} will become proportional to $N$ instead of $N^2$, thus the ``square-law'' doesn't hold any more. 


\subsection{Issue 7: Is RIS Really Intelligent?}
Many existing works \cite{Wu'19,Huang'18'2,LinglongDai,Basar'19} have reported that RISs can enhance wireless communications by properly adjusting the phase shifts of their elements, which has sufficiently shown the reconfigurable characteristic of RISs. However, as the full name of RIS suggests, RISs are both {\it reconfigurable} and  {\it intelligent}. Although the  {\it reconfigurability} of RISs has been widely investigated in the literature, the {\it intelligence} of RISs are still worth further investigation.
\par
To advance the expected smart wireless environment, making RISs more intelligent can be an attractive direction. Particularly, empowering RISs with artificial intelligence (AI) is a promising solution to address some challenges particularly introduced by RIS, such as the high overhead for \ac{csi} acquisition \cite{Huchen} and high complexity for real-time beamforming \cite{Huang'18'2}. For examples, deep compressive sensing (DCS) can be utilized to reduce the pilot overhead for RIS channel estimation. The generative adversarial network (GAN) can be used to generate virtual RIS channels to reduce the frequency of \ac{csi} acquisition. Moreover, the deep reinforcement learning (DRL) can be used to generate and dynamically adjust the precoding strategy by observing the receiver performances, which can remove the required modules of \ac{csi} acquisition and beamforming design.

\subsection{Issue 8: How Can RIS Achieve Passive Beamforming Efficiently?}
The key way for RISs to enhance wireless communications is their capability of passive beamforming \cite{Wu'19}, which means that, by properly configuring their elements, RISs can passively reflect signals toward any desired direction. In most existing works, passive beamforming is designed by optimizing the phase-shift matrix according to the acquired \ac{csi}. However, due to the large overhead for CSI acquisition and high complexity of beamforming design, this methodology may be inefficient in practice. The reason is that, to make the reflected link effective \cite{Najafi'20}, the element number of RISs is usually very large (e.g., $N=1600$ \cite{Yang'17}), which introduces high-dimensional channels. It requires a large number of pilots for channel estimations and high complexity for optimizing the RIS phase-shift matrix, which may offset the benefits brought by RISs \cite{Najafi'20}.
\par
In future works, how to realize passive beamforming efficiently is still an important research direction, and one possible solution is the codebook-based method \cite{Jamali'21}. Specifically, the key idea of the codebook-based method is to design a codebook comprising several codewords for beam training. Then, the designed codebook is employed online, and only the codewords are allowed to reconfigure RISs \cite{Jamali'21}. In this way, the passive beamforming design is simplified as a problem of beam selection, while the explicit channel estimation and online phase-shift optimization are not required any more, which makes passive beamforming simpler and more efficient.

\subsection{Issue 9: Is RIS Really Power Saving?}
\begin{figure*}[!t]
	\centering
	\includegraphics[width=0.7\textwidth]{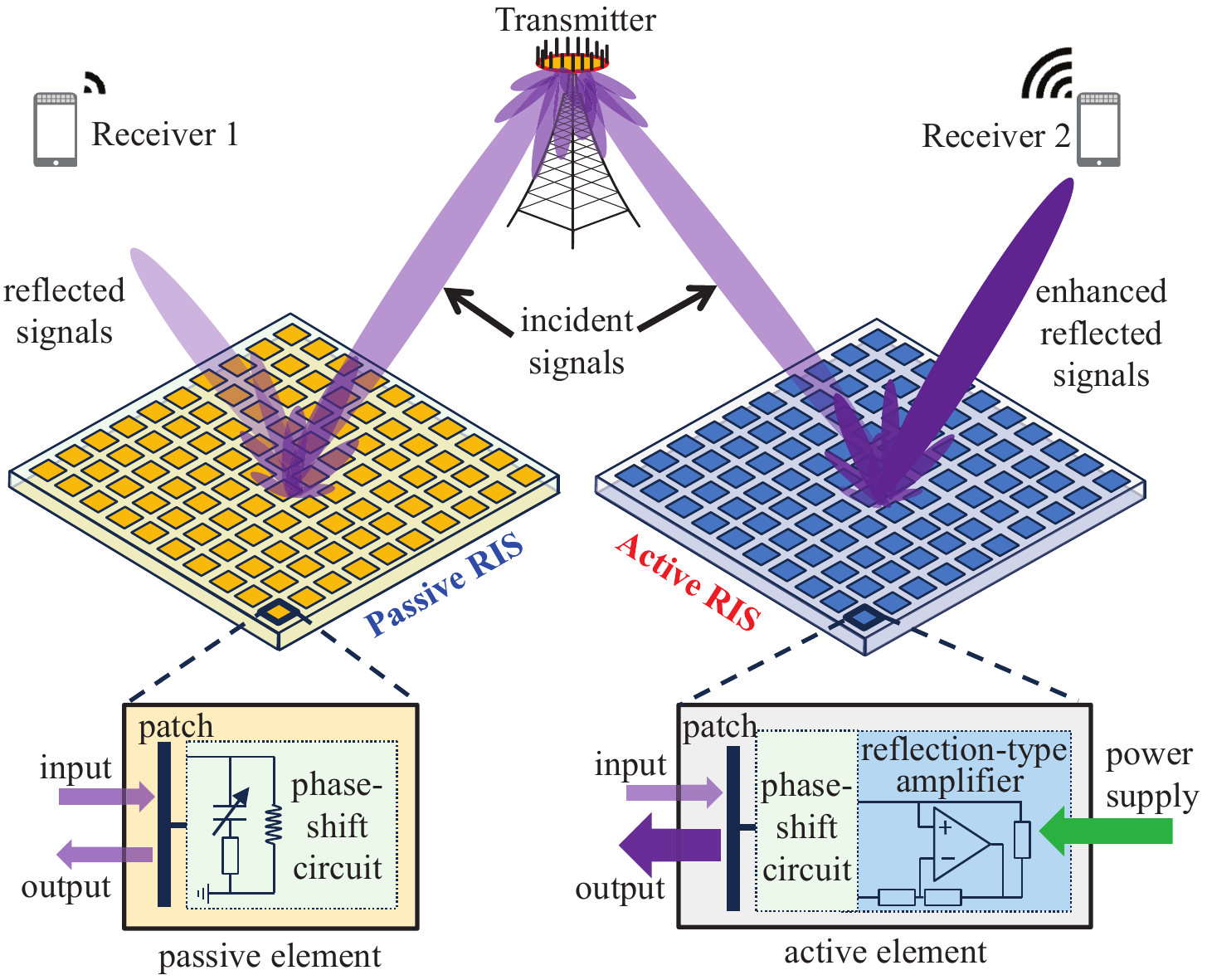}
	\caption{Comparison between passive RIS and active RIS \cite{Zhang'22}.}
	\label{img:activeRIS}
\end{figure*}
Since RISs just passively reflect the incident signals without the need of high-power components such as power amplifier and ADC/DAC \cite{Basar'19}, compared with many existing technologies such as massive MIMO and relay, RISs consume nearly zero power to realize phase shifting \cite{Yang'17}. Therefore, RISs are expected to be able to significantly reduce the power consumption for future 6G communications \cite{Huang'18'2}. However, the practical deployment of RISs may not be energy-saving. The reason is that the static hardware power to drive and reconfigure RISs is required \cite{Huang'18'2}. For example, a typical static power for driving an 1-bit 1600-element RIS is about 11.2 W \cite{Yang'17}. Besides, to reconfigure RISs, additional power is required to support RIS controllers and transmit control signaling. To realize the passive beamforming of RISs, extra power is also required for channel estimation, channel feedback, and beamforming design. As a result, the total power consumption introduced by the deployment of RISs maybe not low. A low-power structure of RISs, as well as the matched transmission methods, are still worth investigating to achieve energy efficiency in future 6G networks.

\section{One Critical Problem and A Potential Solution}
After clarifying nine fundamental issues, in this section, we discuss a critical problem of RISs, i.e., the ``multiplicative fading'' effect. To address this problem, a recently proposed solution called active RIS will be introduced.
\subsection{Critical Problem: ``Multiplicative Fading'' Effect}\label{subsec:fp}
RISs are expected to bring a significant capacity gains  \cite{Wu'19}. However, a critical problem is that, the expected capacity gains can only be achieved in the scenarios where the direct link from the transmitter to the receiver is completely blocked or very weak \cite{Zhang'22}. By contrast, in many scenarios where the direct link is strong, RISs with hundreds of passive elements can only achieve negligible capacity gains. The reason behind this fact is the “multiplicative fading” effect, i.e., in the far-field communications, the equivalent path loss of the reflected link is the product (instead of the summation) of the path loss of the transmitter-RIS link and the path loss of the RIS-receiver link \cite{Wankai'21}. Thus, unless RIS is very close to the transceivers, the equivalent path loss of RIS aided link is usually thousands or even millions of times larger than that of the unobstructed direct link \cite{Najafi'20}. To enable RISs have a noticeable impact on the system capacity, thousands or even millions of RIS elements are required to compensate for this extremely large path loss \cite{Najafi'20}, resulting in unacceptable overhead for channel acquisition and high complexity for real-time beamforming. 
\par
To see this, let us consider an unobstructed SISO system aided by an $N$-element \ac{ris}. Let $\lambda$ denote the wavelength; $d=200$ m, $d_{\rm t}=150$ m, and $d_{\rm r}=200$ m denote the distances between transmitter and receiver, transmitter and \ac{ris}, \ac{ris} and receiver, respectively. Then, according to the “multiplicative fading” model \cite{Najafi'20}, the area of \ac{ris} should be at least scaled as ${\frac{d_{\rm t}d_{\rm r}\lambda}{d}}$ to make the reflected link as strong
as the direct link. Assuming the frequency being 5/10/20 GHz and the half-wavelength element spacing, then $N=10000/20000/40000$ elements will be required. Due to the high overhead of $N$ pilots for channel estimation \cite{Huchen} and the high complexity of ${\cal O}\left(N^2\right)$ for real-time beamforming \cite{Huang'18'2}, such a large number of RIS elements makes the application of RISs in practical wireless networks very challenging. Therefore, to advance the practicability of RISs, a critical issue to be addressed is: \textit{how to overcome the ``multiplicative fading'' effect?}
\subsection{Potential Solution: Active RISs}

\begin{figure}[!t]
	\centering
	\includegraphics[width=0.45\textwidth]{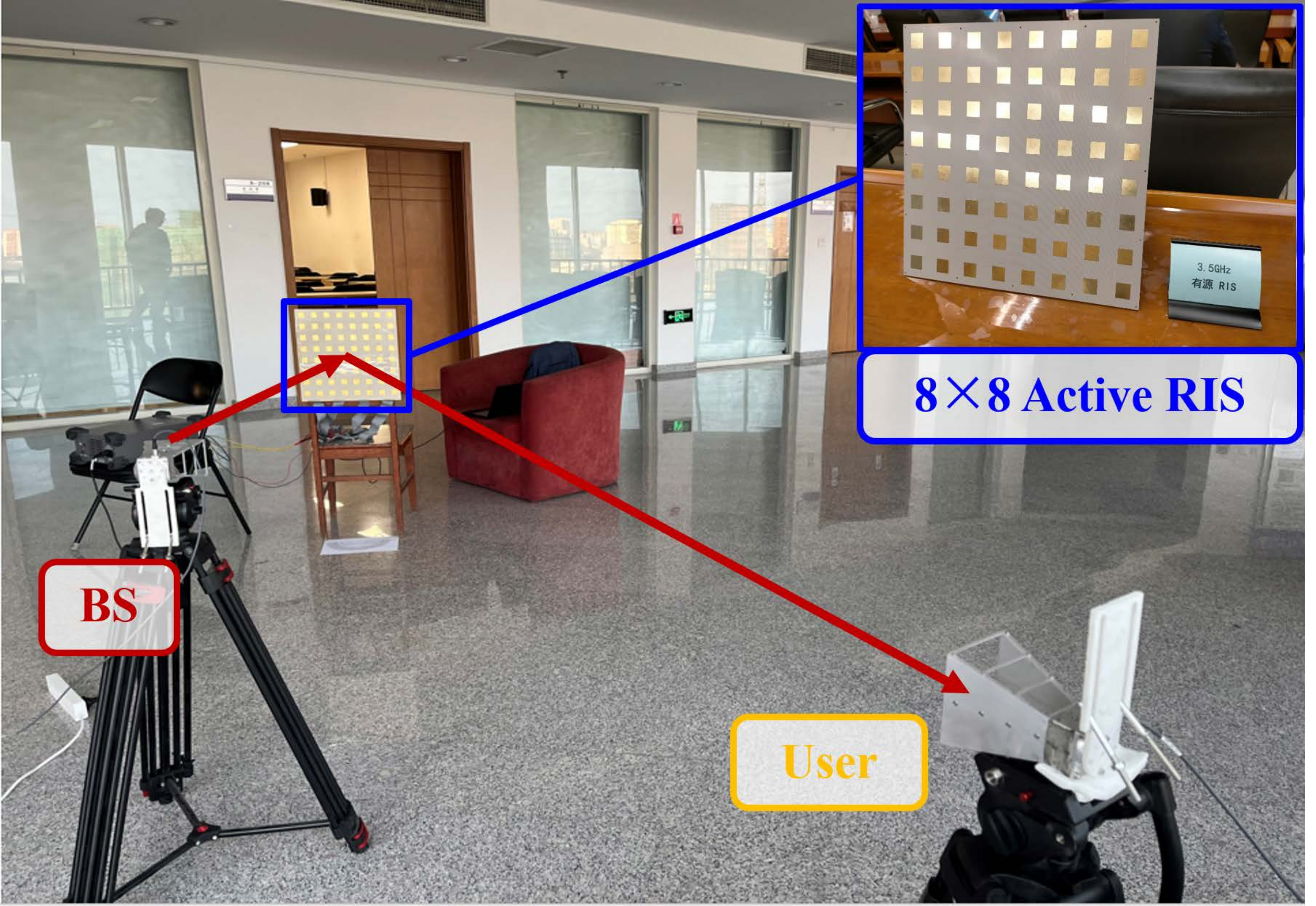}
	\caption{A 64-element active RIS aided wireless communication prototype \cite{Zhan2212:Active}.}
	\label{img:prototype}
\end{figure}

To overcome the ``multiplicative fading'' effect, one promising solution is the active RIS \cite{Zhang'22}. Different from the existing passive RISs, which passively reflect signals without amplification, the key feature of active RISs is to amplify the reflected signals by integrating active components at the RIS elements \cite{Zhang'22}. Specifically, as shown in Fig. \ref{img:activeRIS}, apart from the phase-shift circuit for the passive RIS element, each active RIS element is additionally equipped with an active reflection-type amplifier, which can amplify the reflected signals at the cost of affordable power consumption and hardware cost. Particularly, the reflection-type amplifier can be realized by many low-cost methods, such as current-inverting converter or asymmetric current mirror \cite{Bousquet'12}. Recently, the researchers from Tsinghua University (THU) have developed a 64-element active RIS aided wireless communication prototype working at 3.5 GHz with the bandwidth of 40 MHz, as shown in Fig. \ref{img:prototype}. Based on this prototype, the significant gain of active RISs was validated by experimental measurements \cite{Zhan2212:Active}.

\par
Thanks to the circuit-based hardware structure without RF chains and baseband modules, active RISs still hold many advantages of passive \acp{ris}, such as low cost and low latency \cite{Bousquet'12}. Besides, although active RISs consume additional power to amplify the reflected signals, their hardware power consumption is low. For example, the typical power consumption is 0.123 mW for an amplifier when the reflection gain is set as 22.3 dB \cite{Bousquet'12}. On the other hand, active RISs distinguish passive RISs. Firstly, different from passive RISs which introduce negligible noise, non-negligible thermal noise will be introduced and amplified by active RISs due to the use of active components \cite{Zhang'22}. Secondly, it should be pointed out that, although the additional thermal noise is introduced by active components, active RISs can still achieve the improved \ac{snr} \cite{Changsheng'21}. The reason is that, thanks to beamforming, the desired signals of multiple active RIS elements can be coherently added up with the same phase at the receiver, while the introduced noise cannot. Finally, by using active RISs, the path loss of the reflected link can be equivalently converted from ``multiplicative fading'' to ``additive fading'' \cite{Basar'21}. In other words, active RISs transform the path loss of reflected link from being proportional to $d_{\rm t}^{-2}d_{\rm r}^{-2}$ to being proportional to ${(d_{\rm t}+d_{\rm r})}^{-2}$, where $d_{\rm t}$ and $d_{\rm r}$ denote the distance between transmitter \ac{ris} and that between \ac{ris} and receiver, respectively. This result has proved that, active RISs can indeed overcome the ``multiplicative fading'' effect.

It is worth noting that, due to their different signal models, the signal processing methods for active RISs much differs from those for passive RISs \cite{Zhang'22}. Specifically, passive RISs only reflect signals with controllable phase shifts, which means that only one phase shift matrix is optimized for beamforming. Differently, active RISs integrate reflection-type amplifiers, thus the power allocation among multiple active elements and the power constraint should be considered in the system model \cite{Kunzan'22:CL}. Besides, due to the use of active components, active RISs introduce amplified thermal noise, of which the negative influence should be taken into account while designing the beamforming of active RISs. In addition, since active RISs work in full-duplex mode, their self-interference cannot be simply ignored as passive RISs, which brings challenges for their beamforming design \cite{Zhang'22}. 

\begin{figure}[!t]
	\centering
	\includegraphics[width=0.43\textwidth]{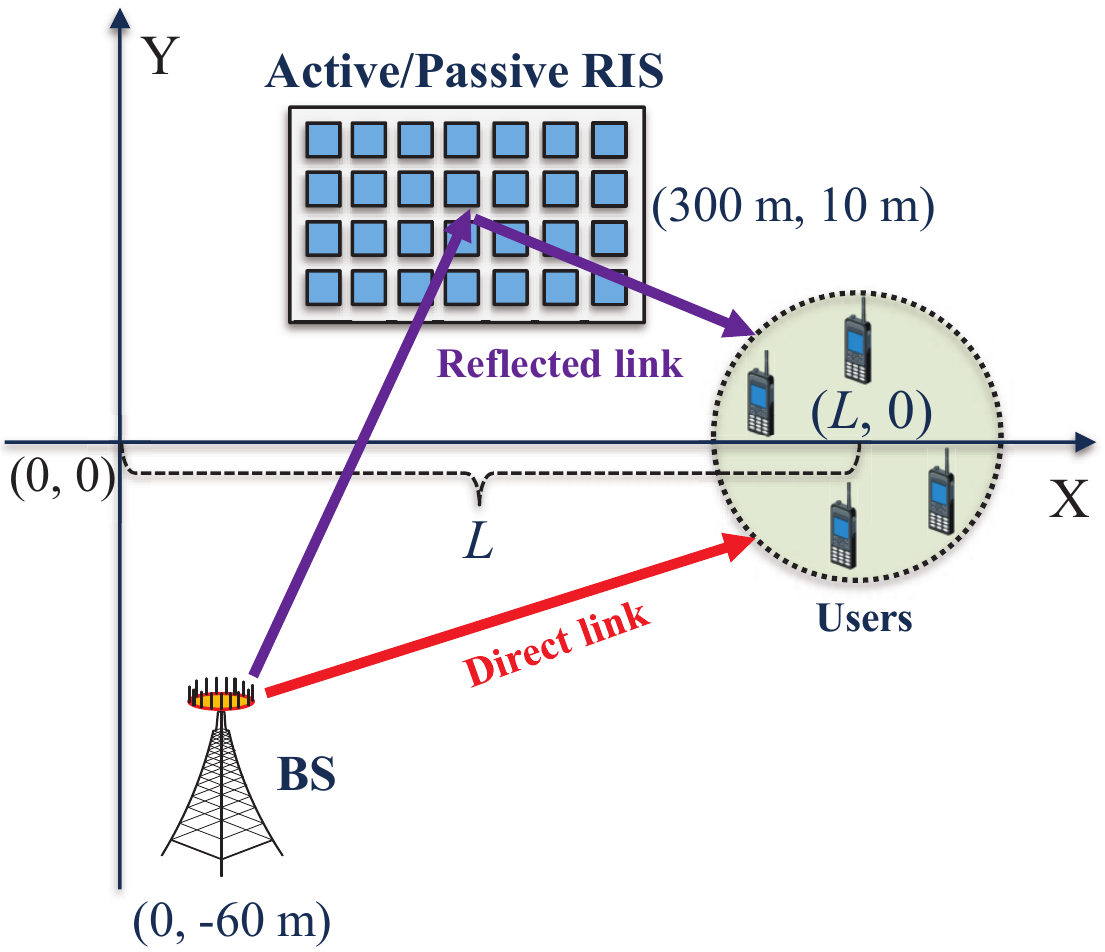}
	\caption{Illustration of the downlink transmission in an active RIS aided MIMO system.}
	\label{img:sim}
\end{figure}
\subsection{Simulation Setup}\label{subsec:sim}

To show the ability of active RISs against the ``multiplicative fading'' effect, we present the simulation supports in this subsection. For simulation setups, we consider the downlink transmission of a MIMO system \cite{Zhang'22}, which includes a 4-antenna \ac{bs}, a 512-element active/passive RIS, and 4 single-antenna users. As shown in Fig. \ref{img:sim}, we assume that the \ac{bs} and the active/passive RIS are located at (0, -60 m) and (300 m, 10 m), respectively. The four users are randomly located in a circle with a radius of 5 m from the center ($L$, 0). The noise power is set to -100 dBm. Let ${P_{\text{BS}}}$ denote the transmit power at the \ac{bs} and ${P_{\text{A}}}$ denote the reflect power of the active \ac{ris}. For fair comparison, we constrain the total power consumption $P:=P_{\text{BS}}+P_{\text A}$ to $10$ dBm by setting $P_{\text{BS}}=P_{\text A}=P/2$ for the active RIS aided system and $P_{\text{BS}}=10$ dBm for the other baselines. We adopt the path loss model ${\rm{PL}} = 37.3 + 22.0\log d$ for three sub-links, where $d$ is the distance between two devices. To account for the small-scale fading, we assume Ricean channel with Ricean factor being one. We consider the following four schemes for simulations\footnote{Simulation codes are provided to reproduce the results presented in this article: http://oa.ee.tsinghua.edu.cn/dailinglong/ publications/publications.html.}:

1) {\it Active RIS:} In an active RIS-aided system, the algorithm proposed in \cite{Zhang'22} is employed to maximize the sum-rate. 

2) {\it Passive RIS:} In a passive RIS-aided system, the algorithm proposed in \cite{Pan'19} is employed to maximize the sum-rate. 

3) {\it Random phase shift:} In a passive RIS-aided system, the RIS phase shifts are randomly set, and then the weighted mean-square error minimization (WMMSE) algorithm \cite{Qingjiang'11} is employed to maximize the sum-rate. 

4) {\it Without RIS:} In a MIMO system without RIS, the WMMSE algorithm \cite{Qingjiang'11} is adopted to maximize the sum-rate.

\subsection{Simulation Result}
\begin{figure}[!t]
	\centering
	\includegraphics[width=0.5\textwidth]{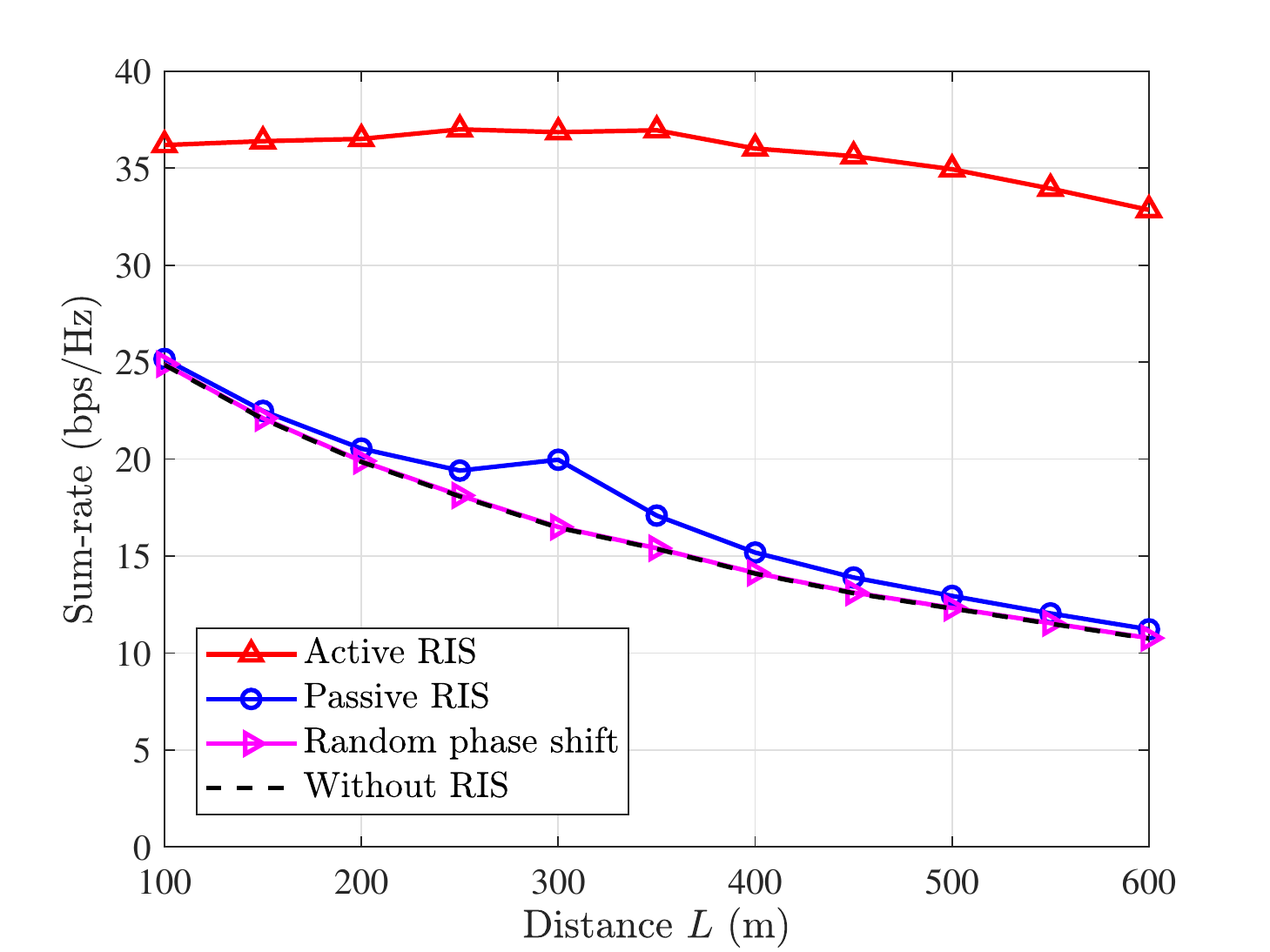}
	\caption{Simulation result of the sum-rate against distance $L$.}
	\label{img:sim_result}
\end{figure}
After averaging the results of 1000 realizations, we plot the sum-rate against distance $L$ in Fig. \ref{img:sim_result}. We can find that, the passive RIS achieves only a limited sum-rate gain, while the active RIS can still realize a noticeable sum-rate gain. For example, when $L=300$ m, the sum-rate without RIS, with passive RIS, and with active RIS are 16.46 bps/Hz, 19.96 bps/Hz, and 36.86 bps/Hz, respectively. For this location, although the users are very close to the RIS, the passive RIS only provides a limited gain of 21\%. The reason is that, due to the ``multiplicative fading'' effect, 512 passive elements can hardly compensate for the severe path loss of the reflected link. It indicates that, the direct link dominates in the equivalent end-to-end link of this scenario, thus the passive RIS has limited contribution to capacity improvement. By contrast, the active RIS achieves noticeable sum-rate gains of 124\%, which is much higher than that achieved by the passive RIS. The reason is that, the active RIS can amplify the reflected signals so as to compensate for the large path loss and make the reflected link effective. These results demonstrate that, compared with the existing passive RISs, active RISs can overcome the “multiplicative fading” effect and achieve noticeable capacity gains in many communication scenarios with strong direct links.

\section{Conclusions and Future Works}
In this article, we have explained the basics of RISs with their design methods, working principles, and potential benefits for future 6G wireless communications. Based on these basics, we have explicitly addressed nine fundamental issues of RISs with highlights on their technical features and key differences from existing technologies, which also clarify some misunderstandings in the literature. Then, one critical problem of RISs has been revealed that, limited by the severe path loss in the reflected link caused by the ``multiplicative fading'' effect, the common RISs can hardly achieve noticeable capacity gains in communication scenarios with strong direct links. To break this fundamental physical limit, a recently proposed solution called active RISs has been introduced. Simulation results demonstrate that, compared with the existing passive RISs, active RISs can overcome the ``multiplicative fading'' effect with much higher capacity gains.

Compared with the traditional passive RISs, active RISs are expected to achieve wider signal coverage, improved capacity, and higher energy efficiency. Some open problems are still worth further investigations. For example, some important performance metrics such as the BS transmit power \cite{Wu'19}, energy efficiency \cite{Huang'18'2}, and user fairness are left for future works. Some potential extensions, such as the applications of active RISs in sensing \cite{Zhao'22}, radar detection \cite{He'22}, localization \cite{Zhang'22:localization}, and mobile edge computing \cite{Hu'21}, are still blank in the literature, which are left for follow-up works.

\balance
\bibliographystyle{IEEEtran}
\bibliography{IEEEabrv,reference}

\begin{IEEEbiography}[{\includegraphics[width=1in,height=1.25in,clip,keepaspectratio]{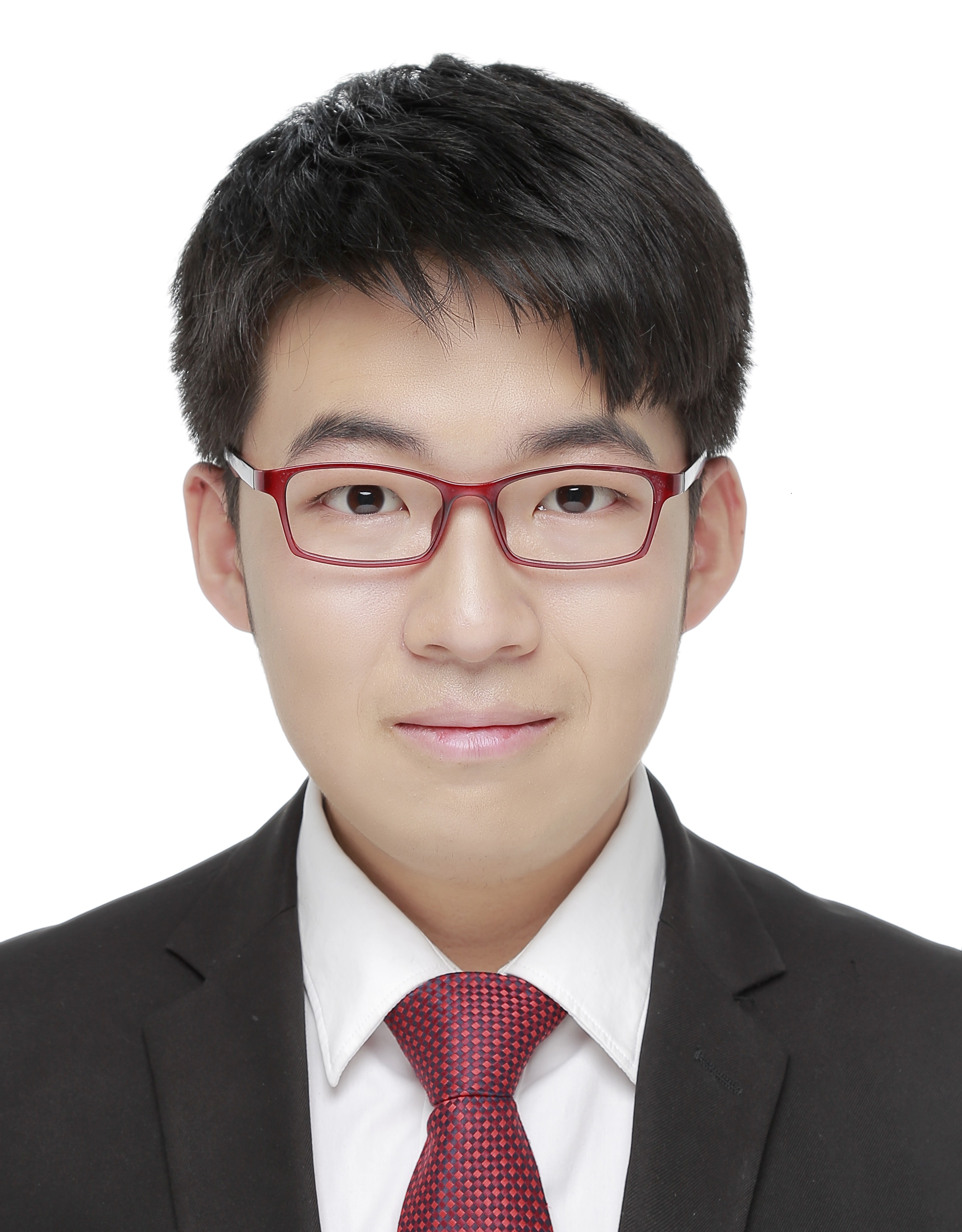}}]{Zijian Zhang}
	(Student Member, IEEE) received the B.E. degree in electronic engineering from Tsinghua University, Beijing, China, in 2020. He is currently working toward the Ph.D. degree in electronic engineering from Tsinghua University, Beijing, China.
	His research interests include physical-layer algorithms for massive MIMO and reconfigurable intelligent surfaces (RIS). He has received the National Scholarship in 2019 and the Excellent Thesis Award of Tsinghua University in 2020.
\end{IEEEbiography}

\begin{IEEEbiography}[{\includegraphics[width=1in,height=1.25in,clip,keepaspectratio]{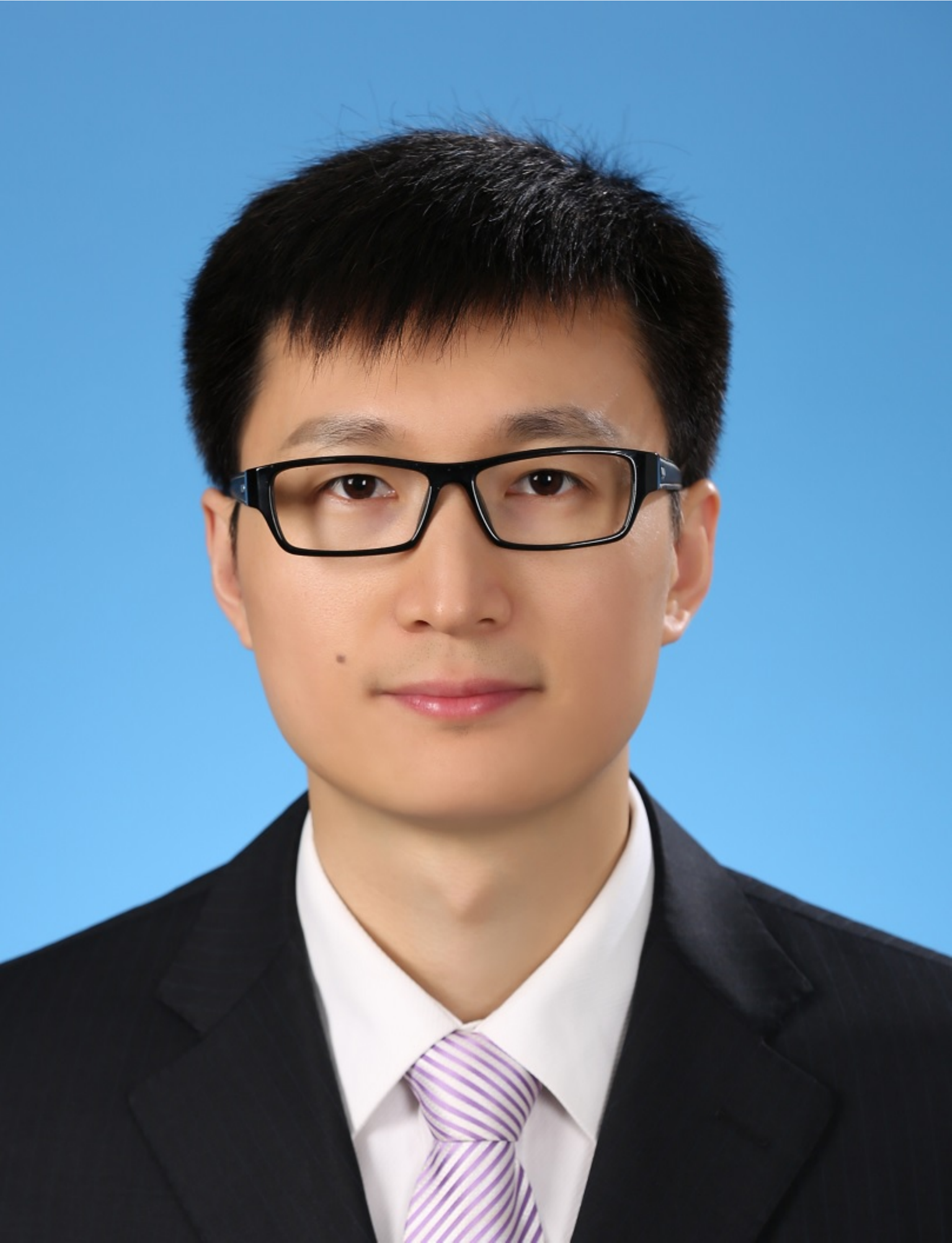}}]{Linglong Dai} (Fellow, IEEE) received the B.S. degree from Zhejiang University, Hangzhou, China, in 2003, the M.S. degree from the China Academy of Telecommunications Technology, Beijing, China, in 2006, and the Ph.D. degree from Tsinghua University, Beijing, in 2011. From 2011 to 2013, he was a Post-Doctoral Researcher with the Department of Electronic Engineering, Tsinghua University, where he was an Assistant Professor from 2013 to 2016, an Associate Professor from 2016 to 2022, and has been a Professor since 2022. His current research interests include massive MIMO, reconfigurable intelligent surface (RIS), millimeter-wave and Terahertz communications, wireless AI, and electromagnetic information theory. He has received the National Natural Science Foundation of China for Outstanding Young Scholars in 2017, the IEEE ComSoc Leonard G. Abraham Prize in 2020, the IEEE ComSoc Stephen O. Rice Prize in 2022, and the IEEE ICC Outstanding Demo Award in 2022. He was elevated as an IEEE Fellow in 2021.
\end{IEEEbiography}

\end{document}